\documentclass[runningheads]{llncs}

\usepackage{amsmath}
\usepackage{verbatim}
\usepackage{amssymb}
\usepackage{graphicx}
\usepackage{url}
\usepackage{booktabs}
\usepackage{multirow}
\usepackage{cleveref}
\usepackage{algorithm}
\usepackage{algorithmic}
\usepackage{xcolor}

\usepackage{blindtext}

\DeclareMathOperator{\arccosh}{arccosh}

\title{HEAT: Hyperbolic Embedding of Attributed Networks}

\author{David McDonald\inst{1}\orcidID{0000-0002-0540-8254}\and
Shan He\inst{1}}
\institute{The University of Birmingham, Edgebaston, Birmingham, B15 2TT, UK
\email{\{dxm237,s.he\}@cs.bham.ac.uk} }

\begin{document}
	
\maketitle

\begin{abstract}
Finding a low dimensional representation of hierarchical, structured data described by a network remains a challenging problem in the machine learning community. 
An emerging approach is embedding these \textit{networks} into hyperbolic space because it can naturally represent a network's hierarchical structure.
However, existing hyperbolic embedding approaches cannot deal with attributed networks, in which nodes are annotated with additional attributes. 
These attributes might provide additional proximity information to constrain the representations of the nodes, which is important to learn high quality hyperbolic embeddings.
To fill this gap, we introduce HEAT (Hyperbolic Embedding of ATributed networks), the first method for embedding attributed networks to a hyperbolic space.
HEAT consists of 1) a modified random walk algorithm to obtain training samples that capture both topological and attribute similarity; and 2) a learning algorithm for learning hyperboloid embeddings from the obtained training samples.
We show that by leveraging node attributes, HEAT can outperform a state-of-the-art Hyperbolic embedding algorithm on several downstream tasks.
As a general embedding method, HEAT opens the door to hyperbolic manifold learning on a wide range of attributed and unattributed networks. 
\end{abstract}

\section{Introduction}
The success of machine learning algorithms often depends upon data representation \cite{bengio2013representation}.
Representation learning -- where we learn alternative (low dimensional) representations of data -- has become common for processing information on non-Euclidean domains, such as the domain of nodes and edges that comprise complex networks.  
Prediction over nodes and edges, for example, requires careful feature engineering \cite{grover2016node2vec} and representation learning leads to the extraction of features from a graph that are most useful for downstream tasks, without careful design or a-priori knowledge.

An emerging representation learning approach for complex networks is hyperbolic embedding. 
This approach is based on compelling evidence that the underlying metric space of many complex networks is hyperbolic \cite{krioukov2010hyperbolic,alanis2016efficient,alanis2016manifold} .  
A hyperbolic space can be interpreted as a continuous representation of a discrete tree structure that captures the hierarchical organisation of elements within a complex system \cite{krioukov2010hyperbolic}. 
Since real world networks are often hierarchical -- where nodes are grouped hierarchically into groups in an approximate tree structure \cite{papadopoulos2011popularity}. 
Hyperbolic spaces therefore offer a natural continuous representations of hierarchical complex networks \cite{krioukov2009curvature,krioukov2010hyperbolic}. 

However, existing hyperbolic embedding approaches cannot deal with attributed networks, of which nodes (entities) are richly annotated with attributes \cite{gibert2012graph,liao2018attributed}. 
For example, a paper within a citation network may be annotated with the presence of keywords, and the people in a social network might have additional information such as interests, hobbies, and place of work. 
These attributes might provide additional proximity information to constrain the representations of the nodes. Therefore, incorporating this node attributes can improve the quality of the final embedding, with respect to many different downstream tasks \cite{liao2018attributed}.

This paper proposes the first hyperbolic embedding method for attributed networks called HEAT.
The method uses a modified random walk algorithm to obtain training samples that capture both topological and attribute similarities.  Inspired by the concept of \textit{homophily} -- that similar nodes are more likely to interact \cite{liao2018attributed} -- the random walk algorithm uses a teleport procedure to capture ``phantom links'' between nodes that do not share a topological link, but have highly similar attributes. 
We then further formulate a learning algorithm to learn the embeddings from the training samples based on stochastic gradient descent in hyperboliod space.

Our results show that, by leveraging node attributes, HEAT can achieve better performance on several downstream tasks compared with a state-of-the-art hyperbolic embedding algorithm.  As a general framework, HEAT can embed both unattributed and attributed networks with continuous and discrete attributes, which opens the door to hyperbolic manifold learning for a wide range of complex networks. 

\subsection{Related Work}
Many hyperbolic embedding algorithms are based on two-dimensional hyperbolic disk, which was popularised by the Popularity-Similarity (or PS) model \cite{papadopoulos2011popularity}.  In  \cite{papadopoulos2011popularity}, Maximum Likelihood (ML) was used in to search the space of all PS models with similar structural properties as the observed network, to find the one that fit it best. This was extended by the authors in \cite{papadopoulos2015network}. 
Due to the computationally demanding task of maximum likelihood estimation, often heuristic methods are used. For example, \cite{alanis2016efficient} used Laplacian Eigenmaps to estimate the angular coordinates of nodes in the PS model. 
The authors then combined both approaches to leverage the performance of ML estimation against the efficiency of heuristic search with a user controlled parameter in \cite{alanis2016manifold}. 
Additionally, \cite{thomas2016machine} propose the use of classical manifold learning techniques in the PS model setting with a framework that they call \textit{coalescent embedding}. 

Beyond the two-dimensional hyperbolic disk, an $n$-dimensional Poincar\'e ball can give more degrees of freedom to the embedding and capture further dimensions of attractiveness than just ``popularity'' and ``similarity'' \cite{nickel2017poincare,chamberlain2017neural}. 
By embedding general graphs to trees, \cite{de2018representation} were able to achieve state-of-the-art results by extending the work of \cite{sarkar2011low}.


		
\section{Background: Hyperbolic Geometry}
In this section, we give some background knowledge about Hyperbolic Geometry to make our paper self-contained. 

\subsection{Models of Hyperbolic Geometry}
Hyperbolic geometry is what one obtains by negating Euclid's fifth postulate, the \textit{parallel postulate}.
In hyperbolic geometry there exists a line $ l$ and a point $ P$ not on $ l$ such that at least two distinct lines parallel to $ l$ pass through $P$ \cite{krioukov2010hyperbolic}.
It has a negative Gaussian curvature $K$ for the entire space and that implies that every single point in the Hyperbolic space is a \textit{saddle point}.
Furthermore, hyperbolic spaces cannot be embedded into a Euclidean space without distortion.
\cite{krioukov2010hyperbolic} informally explain hyperbolic spaces are ``larger'' and have more ``space'' than Euclidean spaces.
This is mathematically reflected by the fact that the area of a circle of radius $r$, does not grow quadratically with $r$, as we are used to in Euclidean space\footnote{Recall that the area of a circle is given by $A=\pi r^2$ in Euclidean geometry.}, but grows exponentially with $A=2\pi \cosh(\zeta r - 1)$.\footnote{$\zeta = \sqrt{-K}$ and $\cosh$ is Hyperbolic cosine function:
$\cosh (x)= (e^{x}+e^{-x}) / 2$}
Because of this property, we observe an exponential expansion of space \cite{boguna2010sustaining}.


Because  of  the  fundamental  difficulties  in  representing spaces of constant negative curvature as subsets of Euclidean spaces,  there  are  not  one  but  many  equivalent  models  of hyperbolic spaces. 
We say the models are equivalent because all models of hyperbolic geometry can be freely mapped to each other by an \textit{isometry}.
Each model (the most popular of which is the \textit{Poincar\'e ball} \cite{krioukov2010hyperbolic,nickel2017poincare} ) emphasizes different aspects of hyperbolic geometry, but no model simultaneously represents all of its properties.
We are free to chose the model that best fits our need.
The most popular model in the network embedding literature is the \textit{Poincar\'e disk} (or \textit{Poincar\'e ball} for $n>2$ dimensions.)
Here, we have the entire hyperbolic plane represented as the interior of a Euclidean unit ball \cite{krioukov2010hyperbolic,nickel2017poincare}.
, sitting in a Euclidean ambient space, where the boundary of the ball represents infinity \cite{krioukov2010hyperbolic,nickel2017poincare}.

The boundary of the ball represents infinity in the hyperbolic system that it is modelling.
Euclidean and hyperbolic distances, $r_e$ and $r_h$ from  the  disk  centre,  or  the  origin  of  the	hyperbolic plane, are related by $r_e =  \tanh(r_h / 2)$
and the shortest path between points (\textit{geodesics}) are given by the diameters of the circle or Euclidean circle arcs that intersect the boundary of the ball perpendicularly.
This model has the main advantage that it is \textit{conformal}: the Euclidean angles in the model are equal to the hyperbolic angles in hyperbolic geometry that the model is representing.
This makes the model a popular choice for embedding methods that abstract node similarity as the angular distance between them, for example: \cite{alanis2016efficient,alanis2016manifold}.
We have also the much less popular \textit{Klein} model of hyperbolic geometry.
This model also represents hyperbolic geometry as a unit disk (or ball) in an ambient Euclidean space.
This model preserves straight lines: straight Euclidean lines map to straight hyperbolic lines.
However, unlike the \textit{Poincar\'e ball}, the \textit{Klein} model is not conformal, and the Euclidean angles in the model are not equal to hyperbolic angles. 


\subsection{The Hyperboloid Model of Hyperbolic Space}
We consider the \textit{Hyperboloid} model primarily in this work.
Unlike disk models, that sit in an ambient Euclidean space of dimension $n$, the Hyperboloid model of $n$-dimensional hyperbolic geometry sits in $n+1$-dimensional Minkowski space-time.
Furthermore, the set of hyperboloid points do not form a disk in this ambient space, but in $n$-dimensional hyperboloid.
We view both the \textit{Poincar\'e} and \textit{Klein} models as (stereographic and orthographic) projections of the points from the hyperboloid to disks orthogonal to the main axis of the hyperboloid \cite{krioukov2010hyperbolic}.
Informally, we can see this relationships as analogous to the relationship between a projected map and a globe \cite{reynolds1993hyperbolic}.

Specifically, $n+1$-dimensional Minkowski spacetime is defined as the combination of $n$-dimensional Euclidean space with an additional time co-ordinate $t$.
Commonly, we denote this set $\mathbb{R}^{n:1}$.
We say that point $\textbf{x}\in \mathbb{R}^{n:1}$ has time co-ordinate  $x_i^0$ and spacial coordinates $x_i^k$ for $k=1,2,...,n$.

The \textit{Minkowski bilinear form} is defined as 
\begin{align*}
\langle\textbf{x}_i, \textbf{x}_j\rangle_{\mathbb{R}^{n:1}} = -\psi ^2 x_i^0 x_j^0 + \sum_{k=1}^n x_i^k x_j^k 
\end{align*}
where $\psi$ is the speed of information flow in our system (normally set to 1 for simplified calculations). Further details have been omitted for brevity, however the reader is directed to \cite{clough2017embedding} for more details. This bilinear form functions as an inner product (like the Euclidean dot product) and allows use to compute norms in a familiar way:
$
||\textbf{x}||_{\mathbb{R}^{n:1}}:= \sqrt{\langle \textbf{x}, \textbf{x} \rangle_{\mathbb{R}^{n:1}}}
$.
However, it is possible for $\langle \textbf{x}, \textbf{x} \rangle_{\mathbb{R}^{n:1}} < 0$ and so norms may be imaginary.

In fact, the points $\textbf{x}$ satisfying $\langle \textbf{x}, \textbf{x} \rangle_{\mathbb{R}^{n:1}} < 0$ are of particular relevance to hyperbolic geometry as the $n$-dimensional hyperboloid $\mathbb{H}^n$ is comprised of just such points:
$
\mathbb{H}^n = \{\textbf{x}\in \mathbb{R}^{n:1} \mid \langle \textbf{x},\textbf{x}\rangle_{\mathbb{R}^{n:1}} = -1, x_{0} > 0\} 
$.
The first condition defines a hyperbola of two sheets, and the second one selects the top sheet.
Shortest paths (\textit{geodesics}) between points on the model are given by the hyperbola formed by the intersection of $\mathbb{H}^n$ and the two dimensional plane containing the origin and both of the points.

The distance (along the geodesic) between two points $\textbf{x}_u,\textbf{x}_v \in \mathbb{H}^n $ is given by
$
d_{\mathbb{H}^n}(\textbf{x}_u, \textbf{x}_v) = \arccosh(-\langle \textbf{x}_u, \textbf{x}_v \rangle_{\mathbb{R}^{n:1}})
$
and is analogous to the length of the great circle connecting two points in spherical geometry\footnote{The proof for the distance formula is given in \cite{reynolds1993hyperbolic}}.
We also define the tangent space of a point $\textbf{x}\in \mathbb{H}^n$ as 
$
T_\textbf{x} \mathbb{H}^n = \{\textbf{x}' \in \mathbb{R}^{n:1} \mid \langle \textbf{x}, \textbf{x}' \rangle_{\mathbb{R}^{n:1}} = 0\}
$.
We see that $T_\textbf{x} \mathbb{H}^n$ is the collection of all points in $\mathbb{R}^{n:1}$ that are orthogonal to $p$.
It can be shown (in \cite{reynolds1993hyperbolic}) that $\langle \textbf{x}', \textbf{x}'\rangle_{\mathbb{R}^{n:1}} > 0$ $\forall \textbf{x}' \in T_\textbf{x} \mathbb{H}^n$  $\forall \textbf{x} \in \mathbb{H}^n$.
In other words, the tangent space of the hyperboloid is positive definite (with respect to the Minkowski bilinear form) for all points on the hyperboloid.
This property actually defines $\mathbb{H}^n$ (equipped with the Minkowski bilinear form) as a Riemannian manifold \cite{reynolds1993hyperbolic}.
Furthermore, we obtain a positive norm for any vector $\textbf{x}' \in T_\textbf{x} \mathbb{H}^n$, allowing us to preform gradient descent.

\section{Method}

\subsection {Problem Definition}
We consider a network of $N$ nodes given by the set $V$ with $|V|=N$.
We use $E$ to denote the set of all interactions between the nodes in our network.
$E = \{(u,v)\} \subseteq V\times V$.
We use the matrix $W\in\mathbb{R}^{N\times N}$ to encode the weights of these interactions, where $W_{uv}$ is the weight of the interaction between actor $u$ and actor $v$.
We have that $W_{u,v} > 0 \iff (u,v)\in E$.
If the network is unweighted then $W_{u,v} = 1$ for all $(u,v) \in E$.
Furthermore, the matrix $X \in \mathbb{R}^{N \times d}$ describes the attributes of each nodes in the network.
These attributes may be discrete or continuous.
We consider the problem of representing a graph given as $\mathbb{G}=(V, E, W, X)$ as set of low-dimensional vectors in the $n$-dimensional hyperboloid $\{\textbf{x}_v \in \mathbb{H}^n \mid v \in V \}$, with $n << \min(N, d)$.
The described problem is unsupervised.

\subsection{HEAT Overview}

Our proposed HEAT consists of two main components: 
\begin{enumerate}
	\item A novel random walk algorithm, which samples the network to obtain training samples that can capture both topological and attributional similarity.
	\item A learning algorithm that learns hyperboloid embeddings from the training samples.
\end{enumerate} The following subsections describe the detailed algorithms.

\subsection{Sample the network using random walks with teleport}
Following previous works \cite{grover2016node2vec,perozzi2014deepwalk}, we modify the random-walk procedure to obtain training samples that capture both topological and attributional similarity. 
The basic idea is that, in additional to standard random walks which capture topological similarity, we use attributional similarity to `teleport' the random walker to the nodes with similar attributes. 

To this end, we define the attributional similarity $Y$ as cosine similarity of the attribute vectors of the nodes. 
We assign a value of 0 similarity to any negative values.
That is 
\begin{align*}
Y_{uv} = \max \left(
\frac{X^T_u X_v}{||X_u||||X_v||} , 0 \right )
\end{align*}
with $||\cdot||$ the Euclidean norm.

We use cosine similarity because it can handle high dimensional data well without making a strong assumption about the data.
We can change the cosine similarity to a more sophisticated and problem dependant measure of pairwise node attributional similarity. 

We then additionally define $\bar{W}$ and $\bar{Y}$ to be the row normalized versions of $W$ and $Y$ respectively, such that each row sums to 1 (zero rows remain as zero rows). 
Therefore, each row in $\bar{W}$ and $\bar{Y}$ is a probability distribution. In particular, the entry $\bar{W}_{u,v}$ encodes the probability of moving from node $u$ to $v$ based on the strength of the topological link between $u$ and $v$, and $\bar{Y}_{u,v}$ likewise encodes the teleport probability based on attributional  similarity.

Beginning from a source node $s$ in the network, we perform a walk with a fixed length $l$ through the network. Each step in the walk from one node to the next is a stochastic process based on both topological structure and similarity of attributes. We define $0\leq\alpha\leq1$ to be a parameter that controls the trade-off between a topological step and an attribute step in the walk. 
Formally, we use $i$ to denote the $i$th node in the walk ($x_0 = s$), and for each step for $i=1,2,...,l$ we sample $\pi _i\sim U(0,1)$ and determine the $i$th node as follows:
\begin{align*}
P(x_i = v\mid x_{i-1} = u) = 
\begin{cases}
\hat{W}_{uv} &\text{if }\pi_i < \alpha,\\
\hat{Y}_{uv} &\text{otherwise.}
\end{cases}
\end{align*}
We follow previous works like node2vec \cite{grover2016node2vec} and Deepwalk \cite{perozzi2014deepwalk}, by considering nodes that appear close together in the same walk to be ``context pairs''. For a source-context pair $(u,v)$, we aim to maximise the probability of observing $v$, given $u$, $P(v\mid u)$.

To build the set of source-context pairs, $D$, we scan across all walks with a sliding window and add pairs of nodes that appear within the window. 
We call this window size the ``context-size'' and it is a user defined parameter (hereafter denoted $c$) that controls the size of a local neighbourhood of a node. 
Previous works show that increasing context size typically improves performance, at some computational cost \cite{grover2016node2vec}. 

\subsubsection{Negative Sampling}
We define the probability of two nodes sharing a connection to be function of their distance in the embedding space. Nodes separated by a small distances share a high degree of similarity and should, therefore have a high probability of connection. Similarly, nodes very far apart in the embedding space should have a low probability of connection. 
We make the common simplifying assumption that a source node and neighbourhood node have a symmetric effect over each other in feature space (ie: $P(v \mid u) = P(u \mid v)$ for all $u,v \in V$)  \cite{grover2016node2vec}.
To this end, we define the symmetric function 
\begin{align*}
\hat{P}(v\mid u) := -d^2_{\mathbb{H}^n}(\textbf{x}_u, \textbf{x}_v)
\end{align*}
to be the un-normalized probability of observing a link between nodes source node $u$ and context node $v$.
We square the distance because this gives us stable gradients
\cite{de2018representation}\footnotetext{For $\textbf{x}, \textbf{x}'\in \mathbb{H}^n$ $\lim\limits_{\textbf{x}'\rightarrow \textbf{x}} \langle \textbf{x}, \textbf{x}' \rangle_{\mathbb{R}^{n:1}} \rightarrow -1$ and $\lim\limits_{x \rightarrow -1} \partial_x \arccosh^2 (-x) \rightarrow 2$.
Contrast this with $\lim\limits_{x \rightarrow -1} \partial_x \arccosh (-x) \rightarrow \infty$}.
We normalize the probability thusly:
\begin{align*}
P(v\mid u) &:= \frac{1}{Z(u)} \exp\left(\hat{P}(v\mid u) / 2\sigma^2\right) \\
Z(u) &:= \sum_{v'\in V} \exp\left(\hat{P}(v'\mid u) / 2\sigma^2\right)
\end{align*}
However, computing the gradient of the partition function $Z(u)$ involves a summation over all nodes $v\in V$, which for large networks, is prohibitively computationally expensive \cite{grover2016node2vec}.
Following previous works, we overcome this limitation through \textit{negative sampling}.
We define the set of negative samples for $u$, $\Gamma(u)$, as the set of $v$ for we we observe no relation with $u$:
\begin{align*}
\Gamma(u) := \{v \mid (u,v) \not\in D \}
\end{align*}
There is no guarantee that the size of these sets is the same for all $u$, so we further define
\begin{align*}
S_m(u, v) := \{ x_i 
\sim \Gamma(u) \mid i = 1,2,...,m  \} \cup \{v\}
\end{align*}
to be a random sample with replacement of size $m$ from the set of negative samples of $u$, according to a noise distribution $P_n$.
Following \cite{grover2016node2vec}, we use $P_n = U ^ \frac{3}{4}$ the unigram distribution raised to the $\frac{3}{4}$ power.
This means that the probability that a node is selected as a negative sample is proportional to its occurrence probability, that we define to be the number of times that it appeared over all the random walks for the network.

\subsection{Hyperboloid Embedding Learning}
We now formulate a loss function $L$ that encourages maximising the probability of observing all positive sample pairs $P(v\mid u)$ $\forall (u, v)\in D$ and minimising the probability of observing all other pairs.
To this end, we define the loss function $L$ for an embedding 
$\Theta = \{\textbf{x}_u\in \mathbb{H}^n  \mid u \in V\}$ 
to be the the mean of negative log-likelihood of observing all the source-context pairs in $D$, against the negative sample noise:
\begin{align*}
L(\Theta) &= -\frac{1}{|D|} \sum_{(u, v) \in D} \log \left[ \frac{\exp\left(-d^2_{\mathbb{H}^n}(\textbf{x}_u, \textbf{x}_v) / 2\sigma^2\right)}{\sum_{v' \in S_m(u, v)} \exp\left(-d^2_{\mathbb{H}^n}(\textbf{x}_u, \textbf{x}_{v'}\right)/2\sigma^2)}\right]
\end{align*}
We observe that minimising $L$ involves maximising $P(v\mid u)$ $\forall (u, v)\in D$ as required.
This encourages source-context pairs to be close together in the embedding space, and $u$ to be embedded far from the noise nodes $v'$ \cite{nickel2017poincare}.

Because we use hyperboloid model for complex network embedding, unlike previous works \cite{nickel2017poincare,de2018representation} that use the \textit{Poincar\'e ball} model and approximate gradients, our gradient computation is simple and exact \cite{wilson2018gradient}. We follow the three step procedure in \cite{wilson2018gradient} to compute gradients $\textbf{x}_u'$.  
The procedure is based on the fact that the loss function $L$ is defined over the whole ambient Minkowski space $\mathbb{R}^{n:1}$, which is further defined over $\mathbb{H}^n \subset \mathbb{R}^{n:1}$. Therefore, for a given point on the hyperboloid $\textbf{x}_u \in \mathbb{H}^n$, we can compute the gradient of $L$ with respect to $\textbf{x}_u$, denoted $\nabla^{\mathbb{H}^n}_{\textbf{x}_u} L \in T_{\textbf{x}_u} \mathbb{H}^n$.
Then to perform gradient descent optimization, we move $\textbf{x}_u$ along $-\nabla^{\mathbb{H}^n}_{\textbf{x}_u} L$ by a small amount $\eta$ to $\textbf{x}_u'\in T_{\textbf{x}_u} \mathbb{H}^n $.
Finally we map $\textbf{x}_u'$ back to $\mathbb{H}^n$ using an exponential mapping.

Specifically, to compute $\nabla^{\mathbb{H}^n}_{\textbf{x}_u} L$, we first compute the gradient with respect to the Minkowski ambient space $\mathbb{R}^{n:1}$ as
\begin{align*}
\nabla_{\textbf{x}_u}^{\mathbb{R}^{n:1}} L &= \Bigg(-\frac{\partial L}{\partial x^0}\bigg|_{\textbf{x}_u}, \frac{\partial L}{\partial x^1}\bigg|_{\textbf{x}_u}, ..., \frac{\partial L}{\partial x^n}\bigg|_{\textbf{x}_u}\Bigg)
\end{align*}
Let $o_{uv} :=-\arccosh (-\langle \textbf{x}_u, \textbf{x}_v\rangle_{\mathbb{R}^{n:1}} )^2 / 2\sigma^2$.
Then
\begin{align}
    \nabla_{\textbf{x}_u}^{\mathbb{R}^{n:1}} L &= \frac{1}{|D|}
    \sum_{\substack{v \\ (u, v)\in D}} 
    \sum_{ v' \in S_m(u,v) }
    (\delta_{vv'} - P(v\mid u)) \cdot \nabla_{\textbf{x}_u}^{\mathbb{R}^{n:1}} o_{uv'}
    \label{ambient1}
\end{align}
and
\begin{align}
    \nabla_{\textbf{x}_u}^{\mathbb{R}^{n:1}} o_{uv} &= \frac{\arccosh \left(-\langle \textbf{x}_u, \textbf{x}_v\rangle_{\mathbb{R}^{n:1}} \right) }{\sigma^2  \sqrt{\langle \textbf{x}_u, \textbf{x}_v\rangle_{\mathbb{R}^{n:1}}^2 - 1}} \cdot \textbf{x}_v
    \label{ambient2}
\end{align}
where $\delta_{vv'}$ is the Kronecker delta function. 
We then use the vector projection formula from Euclidean geometry (replacing the dot product with the Minkowski inner product) to compute the projection of the gradient with the ambient to its component in the tangent space:
\begin{align}
\nabla^{\mathbb{H}^n}_{\textbf{x}_u} L = \nabla^{\mathbb{R}^{n:1}}_{\textbf{x}_u} L + \langle \textbf{x}_u, \nabla^{\mathbb{R}^{n:1}}_{\textbf{x}_u}L \rangle_{\mathbb{R}^{n:1}} \cdot \textbf{x}_u
\label{projection}
\end{align}
Having computed the gradient component in the tangent space of $\textbf{x}_u$, we define the exponential map to take a vector $\textbf{x}_u' \in T_{\textbf{x}_u} \mathbb{H}^n$ to its corresponding point on the hyperboloid:
\begin{align}
\text{Exp}_{\textbf{x}_u}(\textbf{x}_u') = \cosh(||\textbf{x}_u'||) \cdot \textbf{x}_u + \sinh(||\textbf{x}_u'||) \cdot \frac{\textbf{x}_u'}{||\textbf{x}_u'||}  
\label{exponentialMap}
\end{align}
Where $||\cdot||$ denotes the norm with respect to the Minkowski space: $||\textbf{x}|| := \sqrt{\langle \textbf{x}, \textbf{x} \rangle_{\mathbb{R}^{n:1}}}$.
This is analogous to the exponential map in spherical geometry with maps points from the tangent space of a point on the sphere back to the sphere itself.
Finally, the three-step procedure for computing $u'$ with learning rate $\eta$ is:
\begin{enumerate}
	\item Calculate ambient gradient $\nabla_{\textbf{x}_u}^{\mathbb{R}^{n:1}} L$ (\cref{ambient1,ambient2}),
	\item Project $\nabla_{\textbf{x}_u}^{\mathbb{R}^{n:1}} L$ to tangent $\nabla^{\mathbb{H}^n}_{\textbf{x}_u} L$ (\cref{projection}),
	\item Set $\textbf{x}_u = \text{Exp}_{\textbf{x}_u}\left(-\eta \nabla^{\mathbb{H}^n}_{\textbf{x}_u} L\right)$ (\cref{exponentialMap}).
\end{enumerate}
	
\section{Experimental Setup \& Results}

\subsection{Datasets}
We evaluate HEAT on two citation networks (obtained from \cite{bojchevski2017deep}) and one PPI network (obtained from \cite{hamilton2017inductive}).
We select the PPI network to evaluate performance on a different network type than citation.
For the PPI network, \cite{hamilton2017inductive} use positional gene sets, motif gene sets and immunological signatures as features and gene ontology sets as labels. 
Each connected component represents a different tissue sample. 
For our experiments, we select the largest connected component.
Features for all networks were scaled to have a mean of 0 and a standard deviation of 1.
\Cref{tab:statistics} shows the network statistics of these three networks. 
\begin{table}[!htb]
	\centering
	\caption{Network statistics. \textit{Key:} $N$ is the number of nodes, $|E|$ is the number of edges, $d$ is the dimension of node features, $y$ is the number of classes.}
	\label{tab:statistics}
	\begin{tabular}{c c c c c}
		\toprule
		Network & $N$ & $|E|$ & $d$ & $y$ \\ 
		\midrule
		Cora\_ML & 2995 & 8416 & 2879 & 7 \\
		Citeseer & 4230 & 5358 & 2701 & 6 \\
        PPI & 3480 & 54806 & 50 & 121 \\
		\bottomrule
	\end{tabular}
\end{table}

\subsection{Parameter Settings}
\Cref{tab:parameters} shows the parameter settings used for the following experiments. 
For comparison, we used the open-source implementation of the algorithm described by \cite{nickel2017poincare}.
We used default parameters to train their embeddings.
We fix $\alpha=0.2$ for all experiments. Note that HEAT requires less numbers of epochs ($e_{\text{max}}$) and negative samples ($m$). However, as a trade-off, HEAT introduces four more control parameters, but as shown in Section \ref{subsec:alpha}, the performance of HEAT is robust to the setting of those parameters.
\begin{table}[!htb]
	\centering
	\caption{Parameter settings used for experiments. \textit{Key:} $\eta$ is the learning rate, $e_{\text{max}}$ is the number of epochs, $m$ is the number of negative samples for each positive sample, $b$ is the batch size, $c$ is the context size, $s$ is the number of random walks, $l$ is the random walk length, $e_{\text{b}}$ is the number of epochs with reduced learning rate at beginning of training, $\sigma$ is the Gaussian width, $\alpha$ is the topology/attribute trade off parameter.}
    \label{tab:parameters}
	\begin{tabular}{ccccccccccc}
		\toprule
		{} & $\eta$ & $e_{\text{max}}$ & $m$ & $b$ & $c$ & $s$ & $l$ & $e_{\text{b}}$ & $\sigma$ & $\alpha$ \\
    \midrule
		HEAT & .3 & 5 & 10 & 50 & 3 & 10 & 80 & - & 1. & .2\\
		N\&K & 1. & 1500 & 50 & 50 & - & - & - & 20 & - & -\\
		\bottomrule
	\end{tabular}
\end{table}
\subsection{Network Reconstruction \& Link Prediction}
We use network reconstruction experiment to evaluate the embedding's capacity of reflecting the original data.
After training our model to convergence upon the complete information, we compute distances (according the hyperbolic distance) in the embedding space between all pairs of nodes according to both models. 
We assign the true edges in the network positive labels and all other pairs as negatives.
We then rank node pairs by their distance in increasing order and, with a sliding threshold, compute 
the area under the receiver operating characteristic (AUROC) curve.

Following common practice, to evaluate the embeddings' link prediction ability, we randomly select 15\% of the edges in the network and remove them \cite{bojchevski2017deep}.
We randomly select also an equal number of non-edges in the network.
We then train each model on the incomplete network and rank the pairs of nodes based on distance.
We then use the removed edges as the positives and the selected non-edges as negatives and, again, compute AUROC.

\begin{table*}[!htb]
	\centering
	\caption{AUROC scores (to 3 d.p.) for network reconstruction (Recons.) and link prediction (LP) tasks with embedding dimensions of 5, 10, 25, and 50. All scores are averaged over 30 random starting seeds. When $\alpha = 0.2$ the embedding consider node attributes, while $\alpha = 0$ indicates the embedding only consider network topology. Bold scores indicate the best score for each dimension for each task.}
	\label{tab:reconstructionLinkPredictionResults}
	\begin{tabular}{cc|ccc|ccc|ccc}
		 \toprule
		 {} & {} & \multicolumn{3}{c}{Cora\_ML} & \multicolumn{3}{c}{Citeseer} & \multicolumn{3}{c}{PPI} \\
		 Dim		& {}	& \multicolumn{1}{c}{N\&K} 	& \multicolumn{1}{c}{$\alpha=0.0$}	&  \multicolumn{1}{c}{$\alpha=0.2$} & \multicolumn{1}{c}{N\&K} 	& \multicolumn{1}{c}{$\alpha=0.0$}	&  \multicolumn{1}{c}{$\alpha=0.2$} & \multicolumn{1}{c}{N\&K} 	& \multicolumn{1}{c}{$\alpha=0.0$}	&  \multicolumn{1}{c}{$\alpha=0.2$}\\
 		\midrule
		\multirow{2}{*}{5} & Recons. & 
		0.993 & \textbf{0.995} & 0.992 & 
		\textbf{0.999} & 0.999 & 0.998 & 
		\textbf{0.937} & 0.931 & 0.922\\
		{} & LP &
		0.889 & 0.921 & \textbf{0.960} & 
		0.756 & 0.907 & \textbf{0.958} & 
		\textbf{0.910} & 0.898 & 0.887  \\
		\midrule
		\multirow{2}{*}{10} & Recons. & 
		0.994 & 0.996 & \textbf{0.997} & 
		\textbf{1.000} & 0.999 & 0.999 & 
		0.940  & \textbf{0.950} & 0.944 \\
		{} & LP & 
		0.893 & 0.929 & \textbf{0.968} & 
		0.760 & 0.916 & \textbf{0.963} & 
		\textbf{0.913} & 0.909 & 0.901 \\
		\midrule
		\multirow{2}{*}{25} & Recons. 
		& 0.994 & 0.997 & \textbf{0.998} & 
		\textbf{1.000} & 0.999 & 0.999 & 
		0.941 & \textbf{0.961} & 0.959\\
		{} & LP & 
		0.894 & 0.930 & \textbf{0.972} & 
		0.764 & 0.919 & \textbf{0.967} & 
		0.914 & \textbf{0.915} & 0.908 \\
		\midrule
		\multirow{2}{*}{50} & Recons. & 
		0.994 & 0.997 & \textbf{0.998} & 
		\textbf{1.000} & 0.999 & 0.999 & 
		0.941 & \textbf{0.963} & 0.963  \\
		{} & LP & 
		0.893 & 0.931 & \textbf{0.973} & 
		0.761 & 0.919 & \textbf{0.969} & 
		0.914 & \textbf{0.916} & 0.911  \\
        \bottomrule
	\end{tabular}
\end{table*}
Table \ref{tab:reconstructionLinkPredictionResults} shows that, with  node attributes, i.e., when $\alpha = 0.2$,  HEAT outperformed N\&K on all three networks for the link prediction task.

Interestingly, without node attributes, i.e., when $\alpha = 0$, HEAT still outperformed N\&K on the network reconstruction task and link prediction task on the Cora\_ML network as well as the link prediction task for the Citeseer network and both reconstruction and link prediction for higher dimensional embeddings of the PPI network. 
Such results indicate that HEAT can  embed networks without node attributes in hyperbolic space accurately. 

We can also see that, for the Cora\_ML and Citeseer networks, incorporating node attributes, i.e., when $\alpha = 0.2$, improved the AUROC scores significantly for the link prediction task. 
However, for PPI, incorporating node attributes deteriorates the AUROC score, which suggests that for PPI, the homophily hypothesis, i.e., similar nodes in terms of attributes are more likely to interact, might not hold. 
We also observe that, for all three networks, increasing the hyperbolic embedding dimensions only provides marginal improvement.

\subsection{Node Classification}
We make the assumption that nodes are likely to connect to nodes of the same label, and will also display similar attributes.
We train an embedding using complete topology and attribute information.
Note that the embedding is unsupervised, as it is performed with no knowledge of the ground truth labels.
We project the resulting embedding to the Klein model of hyperbolic space, which preserves straight lines \cite{papadopoulos2015network}.
We use an out-of-the-box logistic regression model with the Klein embedding of the network as input features. 
We record micro-F1 and macro-F1 scores by varying the labelled percentage of nodes from 2\% to 10\%.  
For the PPI network, each protein has multiple labels from the Gene Ontology. To evaluate our model in the multi-label case, we adopt a one-vs-all setting, where we train a separate regression model for each label.

\begin{figure}[!htb]
    \centering
	\includegraphics[width=.9\linewidth]{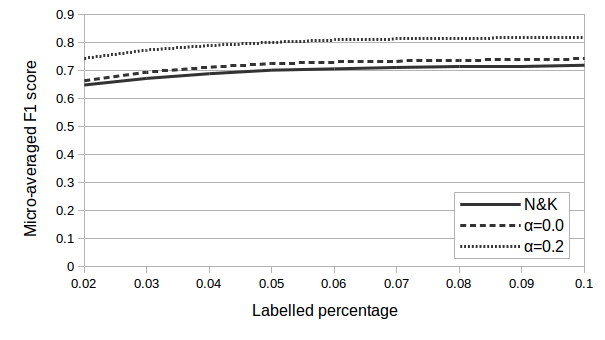}
	\caption{Plots of micro-averaged F1 scores against labelled percentage for the Cora\_ML network. Each score is averaged over 30 random starting seeds. For HEAT, when $\alpha = 0.2$ the embedding considers node attributes, while $\alpha = 0$ indicates the embedding only considers network topology.  We present only the results for an embedding dimension of 10 for clarity, but obtain similar results for all dimensions. }
	\label{fig:coraClassification}
\end{figure}
\begin{figure}[!htb]
    \centering
    \includegraphics[width=.9\linewidth]{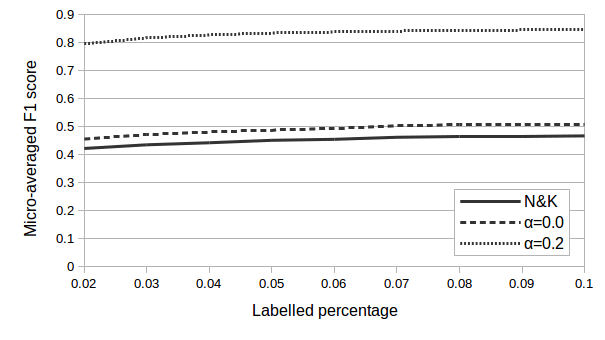}
	\caption{Plots of micro-averaged F1 scores against labelled percentage for the Citeseer network. Each score is averaged over 30 random starting seeds. For HEAT, when $\alpha = 0.2$ the embedding considers node attributes, while $\alpha = 0$ indicates the embedding only considers network topology. We present only the results for an embedding dimension of 10 for clarity, but obtain similar results for all dimensions.  }
    \label{fig:citeseerClassification}
\end{figure}
\begin{figure}[!htb]
    \centering
    \includegraphics[width=.9\linewidth]{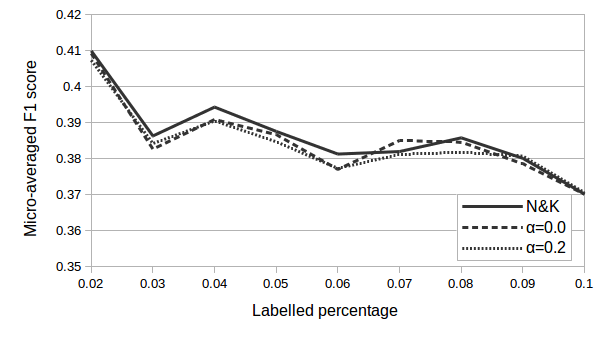}
	\caption{Plots of micro-averaged F1 scores against labelled percentage for the PPI network. Each score is averaged over 30 random starting seeds. For HEAT, when $\alpha = 0.2$ the embedding considers node attributes, while $\alpha = 0$ indicates the embedding only considers network topology. We present only the results for an embedding dimension of 10 for clarity, but obtain similar results for all dimensions. }
    \label{fig:ppiClassification}
\end{figure}
\Cref{fig:coraClassification,fig:citeseerClassification,fig:ppiClassification} 
show that, for node classification, when considering node attributes, i.e., $\alpha=0.2$, HEAT outperformed N\&K on all three networks. We also note that, even without node attributes, HEAT obtained better results on Cora\_ML  and PPI networks than N\&K. 
In contrast, for Citeseer network, without node attributes, HEAT performed much worse than N\&K in terms of micro-average F1 score. These results suggest that, without node attributes, the performance of network embedding algorithms such as N\&K and HEAT depends on the specific network topology.

\subsection{Sensitivity of control parameters} \label{subsec:alpha}
\begin{figure}[!htb]
	\centering
	\begin{minipage}{.9\linewidth}
	\includegraphics[width=\textwidth]{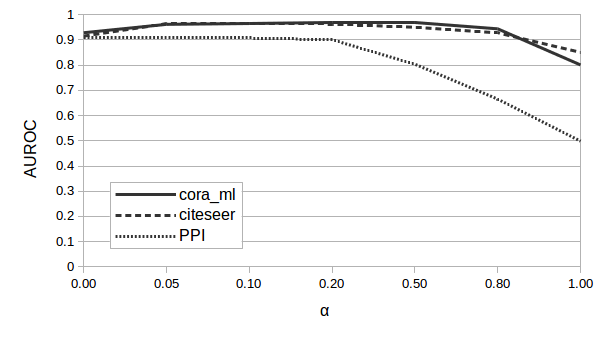}
	\end{minipage}
	\begin{minipage}{.9\linewidth}
	\includegraphics[width=\textwidth]{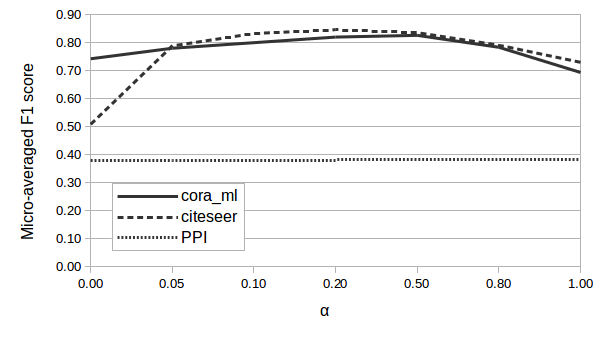}
	\end{minipage}
	\caption{Plots to show the effect of alpha on two downstream machine learning tasks. We fix the embedding dimension to 10 for these experiments. For node classification, we train the logistic regression model with 10\% labelled samples.}
    \label{fig:sensitivity}
\end{figure}

We carried out preliminary experiments to evaluate HEAT's robustness to the setting of the control parameters. Our results indicated that the most sensitive parameter is $\alpha$, which controls the trade off between topology and attributes. We therefore run HEAT over a range of values $\alpha\in\{0.0, 0.05, 0.1, 0.2, 0.5, 0.8,1.0\}$.
\Cref{fig:sensitivity} plot AUROC scores for link prediction and 
the micro averaged F1 scores for node classification obtained from a 10 dimensional embedding with 10\% labelled samples for training the logistic regression model.
From the two figures, we can see the performance of HEAT on the two tasks on the three networks is robust a wide range values of $\alpha$.

\section{Conclusion}
This paper presents HEAT to fill the gap of embedding attributed networks in hyperbolic space. We have designed a random walk algorithm to obtain the training samples that capture both network topological and attributional similarity. We have also derived an algorithm that learns hyperboloid embeddings from the training samples. Our results on three networks show that, by including attributes, HEAT can improve the quality of a learned hyperbolic embedding in a number of downstream machine learning tasks. We also observed that HEAT is competitive compared with N\&K, a state of the art hyperbolic embedding algorithm, on unattributed networks, with smaller numbers of iterations and training samples.

HEAT provides a general hyperbolic embedding method for both unattributed and attributed networks, which opens the door to hyperbolic manifold learning on a wide ranges of networks. 

\section{Code Availability}
Open source software is freely available at \url{https://github.com/DavidMcDonald1993/heat}.

\bibliographystyle{splncs04}
\bibliography{references}

\end{document}